# Quantum Nature of Light Measured With a Single Detector

Gesine A. Steudle[1]*, Stefan Schietinger[1], David Höckel[1], Sander N. Dorenbos[2], Valery Zwiller[2], and Oliver Benson[1]

**The introduction of light quanta by Einstein in 1905 triggered strong efforts to demonstrate the quantum properties of light 'directly', without involving matter quantization. It however took more than seven decades for the 'quantum granularity' of light to be observed in the fluorescence of single atoms[1]. Single atoms emit photons one at a time, this is typically demonstrated with a Hanbury-Brown-Twiss setup[2] where light is split by a beam splitter and sent to two detectors resulting in anticorrelation of detected events. This setup, however, evokes the false impression that a beam splitter is necessary to prove indivisibility of photons. Here we show single-photon statistics from a quantum emitter with only one detector. The superconducting detector we fabricated has a dead time shorter[3] than the coherence time of the emitter. No beam splitter is employed, yet anticorrelations are observed. Our work significantly simplifies a widely used photon-correlation technique[4,5].**

A photon is a single excitation of a mode of the electromagnetic field. The probability $P(n)$ to find exactly $n$ excitations in the mode distinguishes different states of light. Figs. 1a and b show a schematic representation of a coherent state where $P(n)$ is a Poissonian distribution together with a number (or Fock) state with exactly 1 photon per mode, respectively. In the case of a single-photon state ($n$=1) detection of a single excitation projects the measured mode to the vacuum state, i.e. the probability to detect another photon in the very same mode is zero. Since the temporal mode profile is associated with a characteristic coherence time $\tau_c$, coincidence events within the time interval $\tau_c$ are absent, antibunching is observed. On the contrary, for a coherent state the probability to detect a second photon within the same mode is unchanged. Antibunching is thus *not only* a consequence of photons being indivisible particles but requires a specific quantum statistical distribution of discrete excitations. The latter requirement is overlooked in a simple classical explanation of antibunching in a Hanbury Brown and Twiss (HBT) experiment (Fig. 1c). There a photon is regarded as a classical indivisible particle and necessarily has to 'decide' which path to take when impinging on a beam splitter. Such an interpretation is too naïve[6] and even led to paradoxical conclusions, such as Wheeler's delayed choice paradox[7].

Today, many different sources have been realized that generate antibunched light such as single-photon sources based on single atoms[8,9], ions[10], molecules[11,12], colour centres[13], or semiconductor quantum dots[14]. Another approach utilizes quantum correlations between photon pairs to *herald* the presence of a single excitation in a specific mode[15]. Photon statistics is typically measured in the above mentioned HBT setup. However, the *only* reason to use a beam splitter and two detectors is to circumvent the detector's dead time $\tau_d$. For example, commercial avalanche photodiodes (APDs) have dead times of 50 ns - 100 ns or longer, preventing the detection of coincidence events within the coherence time of typical single-photon sources which is on the order of a few nanoseconds. Although more recent experiments could generate single photons with coherence times up to several microseconds[16,17,18] HBT setups are still used.

[1] Humboldt-Universität zu Berlin, AG Nanooptik, Newtonstr. 15, 12489 Berlin, Germany

[2] Kavli Institute of Nanoscience, Delft University of Technology, P.O. Box 5046, 2600 GA Delft, The Netherlands

* email correspondence: steudle@physik.hu-berlin.de

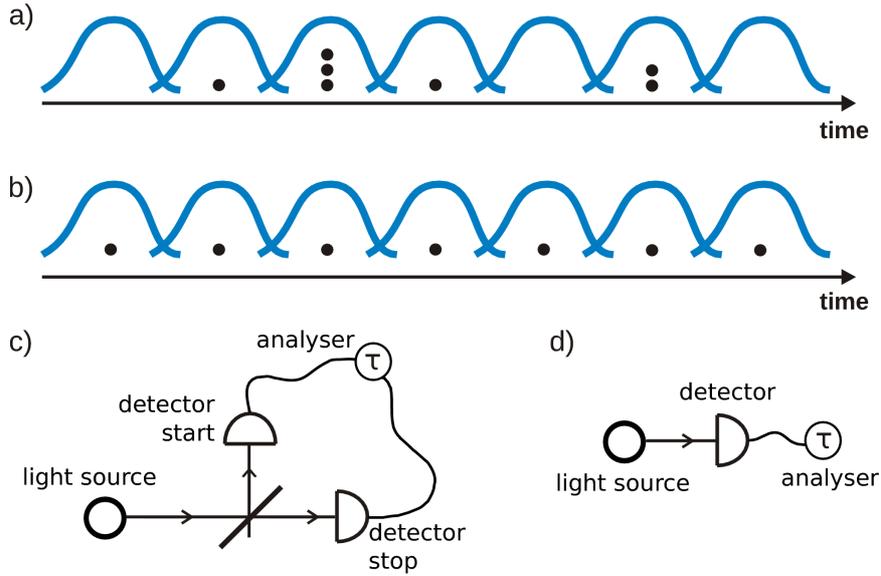

FIG1: **Schematics of quantum states of light and experimental configurations for their statistical analysis**

*Cartoon of a coherent state (a) and a single-photon Fock state (b). Spatial-temporal modes (indicated by blue lines) of a specific coherence time are populated by discrete excitations. For a coherent state this image corresponds to a snapshot, since the number of excitations per mode n can only be predicted with a certain probability P(n). In a single photon Fock state there is exactly one excitation per mode. A detection event projects the mode to the vacuum state. (c) Schematics of a standard Hanbury-Brown and Twiss setup, where the light is split on a beam splitter allowing intensity correlation measurements within time intervals shorter than the individual detector dead times. (d) Direct statistical analysis of light by detecting single photon arrival times with a single detector.*

Recently, measurements of photon statistics with single detection devices based on a gated Geiger mode InGaAs APD[19] or a modified streak camera in single-photon counting mode[20,21] were reported. Non-classical dynamic features were observed in the light from semiconductor microlasers. However, it was so far not possible to detect non-classical properties of light from a single quantum emitter[22] because of poor detection efficiency.

In our approach we determine the statistical properties of a photon stream from a single emitter by detecting the arrival times of individual photons with a single detector (Fig. 1d). For this, the detector's dead time $\tau_d$ has to be shorter than the characteristic correlation time. In the case of weak excitation this correlation time is the Fourier-limited coherence time $\tau_c$ of the photons corresponding to the lifetime $\tau_{life}$ of the excited state, so we require $\tau_d \lesssim \tau_{life}$. In our experiment we use a nitrogen-vacancy (N-V) centre in diamond[23] as a single-photon source together with a superconducting single-photon detector (SSPD).

N-V centres in diamond where a nitrogen atom replaces a carbon atom with an adjacent vacancy in the diamond lattice are the subject of intense research due to their exceptional role as single-photon sources at room temperature. The optical transition in a N-V centre occurs between two spin-triplet states. At least one additional singlet state introduces an 'off-state'. The fluorescence spectrum of a N-V centre is broadened by higher phonon lines, but has a pronounced zero phonon line peak at 638 nm. At room temperature, single-photon emission with count rates up to $10^6$ s$^{-1}$ can be observed[24]. The lifetime of the excited state in N-V centres in diamond nanoparticles is around 30 ns, which is long compared to other single-photon sources[25].

As a single-photon detector we utilized a fibre-coupled SSPD[3] (see **Methods**). It consists of a NbN meandering nanowire coupled to a single-mode fibre (Fig. 2a). Its dead time $\tau_d$ is limited by its kinetic inductance[26], here $\tau_d < 5$ ns. Yet, this value is clearly shorter than the N-V centre's lifetime.

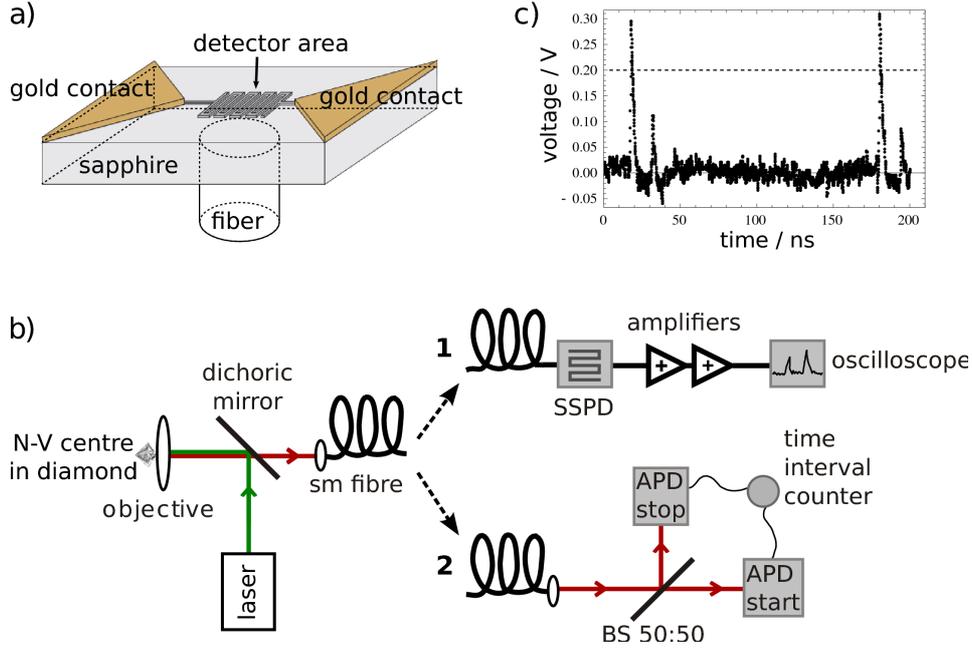

FIG2: **Experimental setup**
(a) Scheme of the fibre-coupled detector chip. (b) A single N-V centre in a diamond nanocrystal is excited with a 532 nm cw laser. Its emission is collected via a confocal optical microscope, coupled to a single mode (sm) optical fibre, and detected in two configurations: 1. a single fibre-coupled SSPD and 2. a standard free beam Hanbury-Brown-Twiss (HBT) setup containing a beam splitter (BS) and two avalanche photodiodes (APDs). Correlations are analyzed by a fast 1 GHz oscilloscope or by a time interval counter. (c) Typical voltage trace with two detection events recorded with the oscilloscope. Only double events with a time difference of between 5 ns and 200 ns were collected.

Antibunching is quantified by measuring the second-order auto-correlation function of the electric field, given by

$$g^{(2)} = \frac{\langle E(t)^\dagger E(t+\tau)^\dagger E(t) E(t+\tau) \rangle}{\langle E(t)^\dagger E(t) \rangle^2} = \frac{\langle : I(t) I(t+\tau) : \rangle}{\langle I(t) \rangle^2}$$

where $E^\dagger$ and $E$ are the field operators and $:\ :$ denotes normal ordering. For uncorrelated light, e.g. laser light, with a Poissonian photon number distribution, $g^{(2)}(\tau) = 1$ for all $\tau$. However, for a number state $|n\rangle$, at $\tau = 0$ it drops to $g^{(2)}(0) = 1 - 1/n < 1$.

Fig. 2b shows the experimental setup for measuring the $g^{(2)}$-function. Fluorescence was coupled into a single-mode fibre, and detection was done in two configurations. Configuration 1 is the single detector setup, i.e. the light was sent via the optical fibre directly to the SSPD. In configuration 2, the HBT setup, light was coupled for comparison into a standard free space HBT setup consisting of a beam splitter and two APDs. In configuration 1, the amplified electrical pulses from the SSPD were fed to an oscilloscope with 1 GHz bandwidth. The oscilloscope was programmed to save a pulse trace whenever a trigger level of 200 mV was exceeded twice with a time difference of between 5 ns and 200 ns (see Fig. 2c). For measuring the $g^{(2)}$-function, 30,000 traces were recorded and analysed. In configuration 2, the time intervals between signals from the two APDs were recorded with a time interval counter.

We first performed two test experiments with classical light. Light from an attenuated laser was coupled into only one of the APDs and the detector clicks were fed into the oscilloscope. The measured $g^{(2)}$-function (Fig. 3a) shows an absence of coincidence counts at time intervals shorter than 30 ns. This is due to the dead time of the APD preventing detection of coincidence events within a 30 ns time interval. It is interesting to note the similarity of this

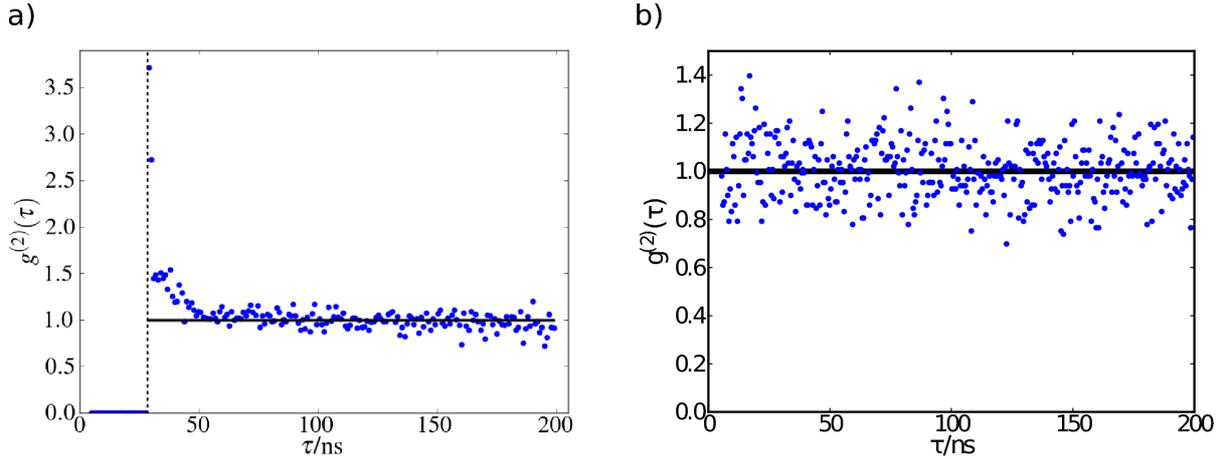

*FIG3: **Correlation measurements with classical light***

***(a)*** *Measured $g^{(2)}$-function of light from an attenuated laser beam sent to a single commercial APD and analysed via the fast oscilloscope by evaluating traces when a second detector event occurs within a time window of 5 ns - 200 ns. The dashed line indicates the dead time of the APD. The bunching observed between correlation times of 30 ns and 50 ns is due to afterpulsing. For a rate of 300,000 s$^{-1}$ and an afterpulsing probability of 0.5 % one out of eleven events is an afterpulsing event and contributes to the area above the solid line ($g^{(2)}(\tau)=1$).* ***(b)*** *Same measurement, but with a SSPD with a dead time below 5 ns. Since only time differences larger than 5 ns were recorded no dead time effect was resolved. The absence of any correlation indicates a Poissonian photon number distribution. The black line is a linear fit.*

classical suppression of coincidences compared to antibunching where the 'suppression' is due to the quantum mechanical projection of a quantum state. For correlation times between 30 ns and 50 ns a bunching feature is observed due to the APD's afterpulsing. The afterpulsing probability according to the manufacturer is 0.5 %. Here this is relevant since at our photon count rates of around 300,000 s$^{-1}$ the probability for a second photon to arrive within a time window between 30 ns and 200 ns after a first one is 5 %, i.e. one out of eleven events when a second pulse is detected is due to afterpulsing. These events account for the bunching observed in Fig. 3a. In a second test we coupled attenuated laser light into the SSPD and again measured the $g^{(2)}$-function (Fig. 3b). Obviously, there are no correlations between incoming photons and, more important, no suppression of coincidence events in the time window of interest (5 ns - 200 ns). Furthermore, no afterpulsing is observed.

Finally, a nanodiamond sample was prepared (see **Methods**), and a single N-V$^-$ centre was located using an inverted microscope as described elsewhere[27]. The 532 nm continuous wave excitation light was filtered out before coupling the emission into a single-mode optical fibre. We measured the $g^{(2)}$-function in the two different configurations of Fig. 2b, the single-detector setup (configuration 1), and the standard HBT setup (configuration 2), respectively. The experimental results are shown in Fig. 4. Blue dots correspond to the $g^{(2)}$-function measured with the single SSPD detector. It has a pronounced antibunching dip which fits well to a three-level rate equation model (solid black line). Red dots correspond to the standard HBT measurement. Obviously, both measurements reveal the quantum nature of the photon stream in the same manner proving that a statistical analysis of a stream of single photons can very well be performed with a single detector only.

In conclusion, our results highlight the fact that there is no need to introduce a beam splitter 'dividing' photons and two detectors for demonstrating the quantum nature of light. On the contrary, our fast detector has such a short dead time that statistical analysis of the detection events yields identical results. Fast detection of single-photon states is highly attractive to study more complex non-classical photon states, such as superpositions of different modes in the temporal rather than in the spectral[28] or spatial[29] domain. Antibunching from sources with shorter coherence times can be measured by further reducing the detector's dead time

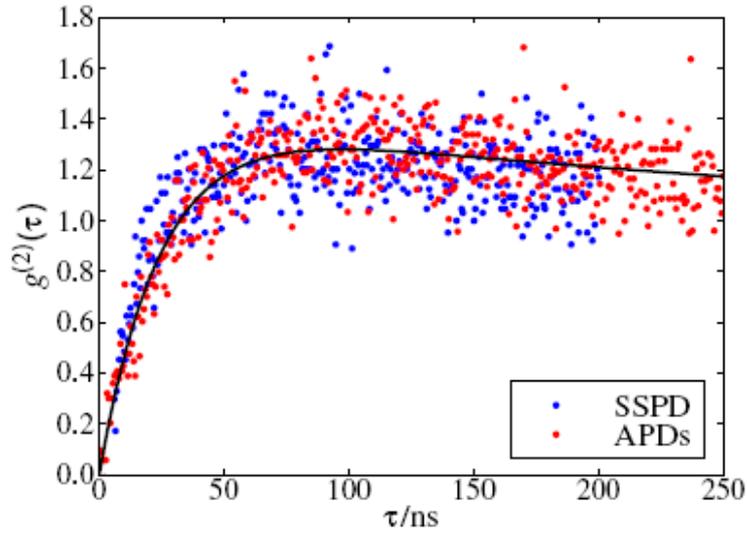

FIG4: **Antibunching measured with a single detector**

*Measurement of the $g^{(2)}$-function in the single detector configuration 1 of Fig. 2b (blue dots) and the standard HBT configuration, configuration 2 of Fig. 2b (red dots), respectively. The black line is a fit to a three-level rate equation model. An additional bunching is observed due to occasional population of a metastable singlet state.*

which is here limited by its kinetic inductance. For example shortening the wire or using a parallel meander geometry are possible routes[30]. Finally, even beyond fundamental considerations such detectors are useful for applications in fluorescence correlation spectroscopy[4] in order to simplify the experimental apparatus.

# METHODS

**Superconducting Detection System.** We use a fibre-coupled superconducting single-photon detector consisting of a 100 nm wide and 5 nm thick NbN meandering wire. The meander structure is fabricated on a sapphire crystal and covers an area of 10 µm x 10 µm. The detector was coupled to an optical fibre. For that purpose the sapphire was thinned to a thickness of 100 µm and its back side was polished. The end facet of a single-mode fibre was aligned on the detector area and glued directly on the back side of the sapphire crystal. The detector chip is mounted on a dip-stick immersed in liquid helium (4.2 K) and biased at 90% of its critical current. The pulses from the detector are amplified by 76 dB with two 2 GHz bandwidth amplifiers and fed to an oscilloscope with 1 GHz bandwidth. Fig. 2c shows a typical response curve recorded with the oscilloscope for two photons absorbed within a time window of 200 ns. There was always a small reflection of the pulse observed after 14 ns which we attribute to impedance mismatch. The dark count rate of the detector was < 50 s$^{-1}$ and its overall efficiency at 630 nm was around 10 %. The temporal resolution of the whole detection system (detector and correlation electronics) was measured by applying femtosecond optical pulses from a titanium-sapphire laser. The voltage pulses of the SSPD and the trigger signal from the laser were connected to a time interval counter with 4 ps resolution. We found a time resolution of our system of 100 ps.

**Sample preparation.** An aqueous suspension of nanodiamonds (diluted approx. 1:50 from the original batch (Microdiamant AG MSY 0-0.05)) is spin-coated onto a glass coverslip. Among the nanodiamonds, only around 1 % contains a single N-V center, deduced from combined optical and atomic force microscopy measurements. Typical heights between 20 and 35 nm are found while the lateral extension is up to twice these values.

# ACKNOWLEDGEMENTS

We acknowledge financial support by the German Funding Agency, DFG (SFB 787), the Federal Ministry of Education and Research, BMBF (KEPHOSI), the Foundation for Fundamental Research on Matter, FOM, and the Netherlands Organisation for Scientific Research, NWO (VIDI).